\documentclass[11pt]{article}
\usepackage[onecolumn]{emulateapj}

  \newcommand {\Ha}     {H$\alpha$}

  \newcommand {\HeII}   {\ion{He}{2}}  
  \newcommand {\OVI}    {\ion{O}{6}}   
  \newcommand {\OV}     {\ion{O}{5}}   
  \newcommand {\OIV}    {\ion{O}{4}}   
  \newcommand {\OIII}   {\ion{O}{3}}   
  \newcommand {\NV}     {\ion{N}{5}}   
  \newcommand {\NII}    {\ion{N}{2}}   
  \newcommand {\NeV}    {\ion{Ne}{5}}  
  \newcommand {\CIV}    {\ion{C}{4}}   
  \newcommand {\CIII}   {\ion{C}{3}}   
  \newcommand {\CII}    {\ion{C}{2}}   
  \newcommand {\SII}    {\ion{S}{2}}   
  \newcommand {\SiIV}   {\ion{Si}{4}}  

  \newcommand {\kms}      {km~s$^{-1}$}

\def\fluxden{\:{\rm ergs\:cm^{-2}\:s^{-1}\:arcsec^{-2}}}

\def\vel{$\rm km~s^{-1}$}

\begin{document}

\title{A Detailed Analysis of a Cygnus Loop Shock-Cloud Interaction}

\author{Charles W. Danforth\altaffilmark{1}, William P. Blair\altaffilmark{1}, John C. Raymond\altaffilmark{2}}

\altaffiltext{1}{Department of Physics and Astronomy, The Johns Hopkins University, 3400 N. Charles Street,  Baltimore, MD 21218; danforth@pha.jhu.edu, wpb@pha.jhu.edu}
\altaffiltext{2}{Harvard-Smithsonian Center for Astrophysics, 60 Garden St., Cambridge, MA 02138; jraymond@cfa.harvard.edu}


\begin{abstract} 

The XA region of the Cygnus Loop is a complex zone of radiative and
nonradiative shocks interacting with interstellar clouds.  We combine
five far ultraviolet spectral observations from the Hopkins Ultraviolet
Telescope (HUT), a grid of 24 IUE spectra and a high-resolution longslit
\Ha\ spectrum to study the spatial emission line variations across the
region.  These spectral data are placed in context using ground-based,
optical emission line images of the region and a far-UV image obtained
by the Ultraviolet Imaging Telescope (UIT).  The presence of
high-ionization ions (\OVI, \NV, \CIV) indicates a shock velocity near
170 \kms\ while other diagnostics indicate $v_{shock}\approx$ 140 \kms.
It is likely that a large range of shock velocities may exist at a
spatial scale smaller than we are able to resolve.  By comparing
\CIV$\lambda$1550, \CIII$\lambda$977 and \CIII]$\lambda$1909, we explore
resonance scattering across the region.  We find that a significant
column depth is present at all positions, including those not near
bright optical/UV filaments.  Analysis of the \OVI\ doublet ratio
suggests an average optical depth of about unity in that ion while flux
measurements of [\ion{Si}{8}]$\lambda$1443 suggest a hot component in
the region at just below $10^6$K.  Given the brightness of the \OVI\
emission and the age of the interaction, we rule out the mixing layer
interpretation of the UV emission.  Furthermore, we formulate a picture
of the XA region as the encounter of the blast wave with a finger of
dense gas protruding inward from the pre-SN cavity.

\end{abstract}



\section{Introduction}

The Cygnus Loop has long been held as the canonical example of a middle-aged supernova remnant (SNR) and a great deal of research has been carried out both on the remnant as a whole and on individual regions.  It is large in angular size (2.8x3.5$\rm ^o$) representing, at an assumed distance of 440 pc (\cite{Blair99}), a spatial extent of 21.5x27 pc.  Partially because of its proximity, the foreground extinction is low (E[B-V] = 0.08; \cite{Fesen82}), making observations across the electromagnetic spectrum possible.  Because of this, the Cygnus Loop provides an ideal laboratory for studying shock physics, gas dynamics, and the structure of the surrounding interstellar medium (ISM).

The Cygnus Loop is thought to be due to a cavity explosion (\cite{Levenson97}), whereby the ionizing radiation and stellar wind of a 15-20 M$_\odot$ star (spectral type B0-B1) excavated a rarefied cavity before exploding as a supernova, perhaps as recently as 5000 years ago.  For most of the lifetime of the remnant, the blast wave has been propagating through this cavity and has only recently begun to interact with the denser cavity wall material, emitting bright visible, UV and X-ray radiation.

Where the blast wave encounters diffuse atomic material we see Balmer-dominated, ``nonradiative'' shocks. The northeast rim, in particular, shows a series of long, smooth, exceedingly thin nonradiative  filaments visible in \Ha\ (\cite{Raymond83}; \cite{Long92}; \cite{Hester94}; \cite{Blair99}),  X-ray emission (\cite{Graham95}; \cite{Levenson97}; \cite{Levenson98}), and FUV emission (\cite{BlairVoy91}; \cite{Danforth00}; hereafter Paper~I). Where the cavity wall is denser either from the remains of pre-existing ISM clouds or compressed cavity gas, the blast wave decelerates quickly, cooling rapidly through emission of UV and optical line radiation and producing regions of bright radiative filaments.  In the northeast and over much of the western limb, the remnant is bent into complex filamentary structures defined by radiative shocks (\cite{Miller74}; \cite{Blair91}; Paper~I). In the southeast, a prominent X-ray filament representing the primary blast wave is observed while only one small knot of optical emission is seen (\cite{Fesen92}, \cite{Graham95}).  This implies that the blast wave has yet to reach the cavity wall boundary over most of the southeastern rim.

While no formal naming convention exists for different parts of such a large object, certain names have been adopted in the literature.  In this paper we investigate in unprecedented detail the triangular region dubbed ``XA'' by \cite{HesterCox86} at the southern end of the bright northeast radiative region NGC~6992.  This region is notable because of its close  proximity to a steep gradient in bright X-ray emission and bright UV/optical emission.  The former arises from fast ($\sim$400~\kms) shocks in lower density regions, while the latter is due to slower ($\sim$150~\kms) shocks in density enhancements, or `clouds' in the  ISM.  Furthermore, the entire X-ray front is indented by several arcminutes from the circularity seen elsewhere in the Cygnus Loop.  This is indicative of strong deceleration and provides a valuable laboratory for studying the interaction of a shock wave with the ISM.

Two outstanding problems regarding the XA region are the existence of X-rays and high-ionization lines in a region of relatively low inferred shock velocities, and the three-dimensional geometry of the region.  By some diagnostics, the shocks present in the XA cloud are not strong enough to produce high ionization states such as \OVI\ (114eV, this paper), [\NeV] (97eV, \cite{Szentgyorgyi00}) and [\ion{Fe}{10}] (262eV, \cite{Ballet89}).  One explanation is that there is a significant range of shock velocities throughout the cloud and our observations do not resolve the different zones.  A competing view is that the cold, dense cloud is being baked in diffuse, 10$^6$K gas and we see thermally excited \OVI\ arising in a mixing layer (\cite{Slavin93}). 

As regards the physical geometry, it seems clear from images alone that a SNR shock has encountered a cloud and is wrapping around it (\cite{HesterCox86}).  Both spherical and ``inclined cigar'' shapes have been suggested.  However, it is possible that the ``cloud'' we see is nothing more than a finger of denser material protruding inward from the cavity wall.

In this paper, we present several new data sets to investigate this complex region. Five positions across the XA cloud were observed with the Hopkins Ultraviolet Telescope (HUT) during the Astro-2 mission.  These observations provide access to the 900 -- 1850 \AA\ region, but lack spatial information within the spectrograph aperture.  Hence, we also  observed the region extensively with the International Ultraviolet Explorer (IUE) to provide  finer spatial information on the lines longward of 1200\AA.  To constrain the velocity structure along the length of the XA region, we analyze a high-dispersion long-slit spectrum in \Ha\ and [\NII].  We examine these data in the  context of optical and UV imaging data of the region.  We describe the imaging and spectral data in \S2.  In \S3, we discuss resonance line scattering and derive information about  the shock velocities observed in this region.  In \S4 we discuss our current picture of the XA region based on these data.  Conclusions are outlined in \S5.


\section{Observations and Reductions}

\subsection{Imaging} 

To establish as broad a wavelength range as possible in which to study the XA region, and to provide information of the small scale morphology of structures in the regions observed spectroscopically, we have obtained and collected images in the visual band, FUV, and X-ray.  

Optical images of the XA region were obtained in \Ha\ and [\OIII]$\lambda$5007 using the Fred Lawrence Whipple Observatory 1.2 m telescope at Mt. Hopkins, AZ, on 1992 October 13 with narrow band filters 22 and 18\AA\ wide (FWHM), respectively.  Both exposures were 600 seconds in duration and have a pixel scale of 0.652 \arcsec\ pixel$^{-1}$.  Bias and flat field corrections were performed using IRAF\footnote{IRAF is distributed by the National Optical Astronomy Observatories (NOAO), which is operated by the Association of Universities for Research in Astronomy, Inc.\ (AURA) under cooperative agreement with the National Science Foundation.}.  These data were taken under less than ideal conditions, with seeing of 1.5'', and no meaningful flux calibration was obtainable.  Even so, these  images provide a wealth of information on the morphology of filaments in the region and permit the spectral data to be placed in context.

Exposures in the far-UV were taken by the Ultraviolet Imaging Telescope (UIT) concurrently with the HUT data described below (cf. \cite{Cornett92}; Paper~I).  UIT was a part of the Astro-1 and Astro-2 missions on the Space Shuttle in 1990 December and 1995 March; technical specifications are given by Stecher et al. (1992, 1997).  We use the data from four overlapping frames from the Astro-2 mission through the medium width FUV ``B5'' filter with a transmission peak at about 1550\AA\ (see Figure 3b of Stecher et al. 1992).  The B5 filter is broad enough to pass several emission lines as well as $\sim$15\%  
of the hydrogen two-photon continuum when it is present (see detailed discussion in Paper~I). The total UIT exposure time in the region around XA was 6030 s.  The frames were aligned, scaled and trimmed to match the two visual band images using packages in IRAF. We see only a handful of stars in the FUV image; thus, we expect stellar contamination  in our spectra to be negligible.

To provide a better understanding of the spatial relationships of emissions in the XA region, a three-color image was constructed with \Ha\ as red, [\OIII] as green and the UV B5-band image in blue.  This image is shown in Figure~1.  The relative levels of the three channels were adjusted arbitrarily so that the peak emission in each band was unsaturated.  We can see that, in general, the front (eastern, left) edges of the shock fronts are prominent in blue (UV emission) while the red \Ha\ of cooler gas is seen more distributed to the west (in the post-shock region).  Filaments tend to be smooth and narrow in B5 and [\OIII] while the \Ha\ morphology is more frothy and turbulent in appearance.

For comparison, we have also obtained the ROSAT-HRI data (0.1--2.4 keV) of the XA region as described by \cite{Levenson97}.  The spatial resolution of these observations is the best to date in this band, but still limited to a $\sim$6'' FWHM beam size.  Nevertheless, we have aligned these data to match the other images and show them in contour form in Figure~2 overlaid on the [\OIII] image.  The image was interpolated to the same pixel scale used in the other images and contoured using IRAF's IMCONTOUR routine using contour levels at 10\%  
of the image maximum.  Alignment is expected to be accurate to a few arc seconds or better.  We see that the X-ray edge coincides with the easternmost UV/optical filaments and matches very well with the bright [\NeV] ridge in this region reported by \cite{Szentgyorgyi00}.  

\subsection{HUT Observations}

Five HUT observations of the XA region were obtained during the Astro-2 mission in 1995 March.  The HUT spectra cover the wavelength range 850-1850\AA\ at 0.5\AA\ pixel$^{-1}$ with $\sim$3\AA\ resolution. Details of the spectrograph and telescope can be found in papers by \cite{Davidsen91}, \cite{Kruk97} and \cite{Kruk99}.  The HUT spectra have excellent sensitivity even in these relatively short exposures, and sufficient spectral resolution to separate most strong line blends.  They also provide crucial coverage to the Lyman limit, and in particular cover the important \OVI$\lambda\lambda$1032,1038 resonance doublet.  However, their major drawback in regions of complicated emission is that no spatial information within the aperture is forthcoming from the Reticon detector.

In the first pointing, the 10\arcsec$\times$56\arcsec\ rectangular aperture was oriented north-south and stepped from the initial position (HUT3 in our nomenclature below) westward across the region in 20'' increments (HUT2, HUT1). In a second pointing, the HUT3 position was again acquired initially, and then the aperture was stepped eastward from center by the same offsets (HUT4, HUT5).  Observational details are listed in Table~1.  Our confidence in the aperture positioning is 1-2\arcsec.  Unfortunately, all pointings were made during  orbital ``day'' (due to target visibility constraints), and contamination by residual atmospheric airglow is significant. The reconstructed aperture positions are indicated in Figure 3a.

The HUT data were reduced from the raw files produced by the HUT calibration pipeline, using an IRAF-based system put together by the HUT team at JHU.  This reduction system propagates statistical errors along with the data.  The line strengths in the resulting spectra were measured with the SPECFIT package in IRAF (Kriss 1994), which can deconvolve blended line complexes based on various user-supplied criteria.  Both line fluxes and uncertainties generated in this way are presented in Table~2.  A sample spectrum is shown in Figure~4.

By combining all five HUT spectra, we have created a very high signal-to-noise FUV spectrum (Figure~5).  With a total exposure time of 2320 seconds, this is probably the highest signal to noise FUV spectrum of a supernova remnant ever taken and shows what we might expect to see from a more distant remnant in similar surroundings such as N49 in the LMC.  Certain very weak spectral features pop out in this spectrum which are otherwise lost in the noise.  For instance, [\ion{Si}{8}]$\lambda\lambda$1440,1445 is seen at a flux of $9\times10^{-17} \fluxden$.  This is a novel feature at very high ionization arising in 10$^6$K X-ray emitting gas and is comparable to optical [\ion{Fe}{10}] and [\ion{Fe}{14}] measurements made by \cite{Ballet89}.  Other faint lines include \ion{S}{4} $\lambda \lambda$1062.6,1073.0 ($1.4\times10^{-16}$), \ion{Ne}{5}] $\lambda$1146.1 ($6\times10^{-17}$), [\ion{Ne}{5}] $\lambda$1574.8 ($5\times10^{-17}$) and \ion{S}{3}] $\lambda$1728.9  ($6\times10^{-17}$).  The fluxes of these faint lines are in units of $\fluxden$ and are uncertain by perhaps 50\%.  
The \ion{Ne}{6}] $\lambda\lambda$999.6,1006.1 lines were not detected, and we take $5\times10^{-17} \fluxden$ as an upper limit.

\subsection{IUE Observations}  

During its lifetime, the IUE satellite observed many locations in the Cygnus Loop.  In its final year of operations, we were extremely fortunate to be able to use this workhorse one last time in support of the HUT observations in the XA region. We performed an extensive grid of IUE observations to map out the variations of UV line intensities  (primarily within the HUT aperture positions) at finer spatial scales than would otherwise have been available.  Archival IUE data of the general XA region were also incorporated into our analysis.

IUE's SWP spectral range of $\sim$1200-1950\AA\ did not include \OVI\ but does extend up past \CIII]$\lambda$1909.  The effective spectral resolution of the filled large aperture was $\sim$10\AA, so line blending in some regions of the spectrum is significant.  The IUE detector allowed spatial resolution along the 20\arcsec\ length of the aperture perpendicular to the dispersion, as well as from position to position.

Our data set consists of a grid of 24 low resolution SWP spectra, the positions of which are shown schematically in Figure~3b.  The large aperture of the SWP camera was 9.07\arcsec$\times$21.65\arcsec\ in size (with rounded corners not shown in the Figure), so with three IUE positions, a HUT aperture could be covered.  The exposures were designed to cover the same regions as those explored with HUT on the Astro-2 mission as well as to fill in the 10\arcsec\ wide space between HUT positions 1 and 2 (which we will call position 1a).  In our nomenclature, each IUE aperture is named after the HUT position it occupies followed by a north (N), middle (M) or south (S) designation.  Seven of the apertures in the grid are oriented approximately east-west instead of north-south as the HUT apertures are.  These were archival data taken during earlier observing campaigns.  There are also four outlying  fields: SWP~48025 and SWP~56022 which lie slightly NE and W of the main block of HUT exposures, respectively, and SWP~26481 and SWP~48030 which occupy relatively ``empty'' regions (at least from the optical perspective) farther to the west and south, respectively, of the main XA cloud.  

We also obtained archival LWP IUE spectra in the same apertures as four SWP spectra.  These observations extend the wavelength coverage to 3300\AA\ and include the interesting [\CII] $\lambda$2325 line.  The SWP and LWP IUE observations are summarized in Table~3.

The spatially resolved `line-by-line' IUE data files were reduced with the APALL task in IRAF.  Backgrounds were subtracted from adjacent areas of the images and the spectra collapsed to one dimension.  Spectra were divided by exposure time and flux calibrated with an inverse sensitivity function supplied by IUEDAC and interpolated to our resolution.

Since there is some spatial resolution available even within the IUE large apertures, we also reduced the data using $\sim$10\arcsec$\times$10\arcsec\ half-apertures.  These half apertures were reduced just like the full apertures but with two spectra extracted from each data set instead of one.  For four apertures (those nearly perpendicular to the HUT slits) we extracted the middle two fourths of the full aperture.  Full apertures were determined to be 14 pixels FWHM in the cross-dispersion direction giving a scale of 1.4\arcsec\ pixel$^{-1}$.  The pixel scale, however, over samples the instrumental resolution of $\sim$4\arcsec.  In theory, we could have subdivided the apertures into even smaller units.  However, because the detected signal decreases rapidly for smaller extractions, we did not subdivide the apertures more finely in our analysis.

Line strengths were determined with the SPLOT task in IRAF and normalized to $\fluxden$.  The results are presented in Table~4 in the same units used in Table~2.  Since the HUT spectra cover such a wide range of conditions, differences between them are subtle.  IUE spectra, on the other hand, show considerable variation across the face of XA.  In the top two panels of Figure~4, we show two IUE spectra from the northern and southern ends of HUT position 5.  The northern spectrum is typical of a lower ionization region of moderate brightness while the southern position shows a higher ionization region.  The bottom panel shows the HUT spectrum for comparison.

Because IUE was not a photon-counting instrument, errors are not easy to quantify.  We have attempted to assess errors by comparing UV line strengths with associated sections of the UV/optical images.  Since we have an image in [\OIII]$\lambda$5007 and spectral fluxes in the strong \OIII]$\lambda$1665 line, these make a good basis for comparison.  The [\OIII] image is not flux-calibrated, so we scale the it such that the global average [\OIII] flux in all the IUE sampled regions is equal to the average, calibrated flux in the \OIII] spectral lines.  We then calculate the difference between line flux and scaled image flux for each aperture and half aperture.  This quantity was dubbed our ``relative error''.  According to the models of \cite{HRH}, the O$^{++}$ flux ratio varies only on the level of $\sim$10\% 
for shock velocities in the range anticipated.

With the FUV image we can carry out a similar procedure.  FUV image fluxes were compared with  that of a composite ``B5 line'' constructed by summing the expected line fluxes over the filter throughput.  (This is discussed more thoroughly in \S3.1.)

There appears to be a weak anticorrelation between flux and ``relative error'' in \OIII] and a stronger anticorrelation in B5.  Higher fluxes correspond to smaller errors, as expected.  In our simple analysis, we will ignore this trend and estimate a flux-independent uncertainty based on the averages of the relative errors and their standard deviations.  Taking the upper 1$\sigma$ values of these numbers and averaging the results from both bands will, we feel, give the best estimate of the uncertainties for the IUE lines.

A good check of our method can be obtained by comparing the errors generated by this method for the HUT lines and those generated by SPECFIT by more rigorous statistical means.  For both bands, the errors generated are of similar magnitude by the two different methods.  Thus we feel justified in quoting uncertainties of 20\%  
for IUE full-aperture, and 30\%  
for IUE half-aperture measurements of strong spectral lines. It is possible that the weaker lines are uncertain to a factor of 50\% or more.


\subsection{Optical Echelle Spectrum}

A high-dispersion, longslit echelle spectrum of the XA region was obtained by one of us (WPB) in 1992 September, using the KPNO 4m and Cassegrain echelle spectrograph with the UV camera. Grating 79-67\arcdeg\ was used with a flat mirror in place of the cross disperser, and a 73\AA\ FWHM \Ha\ filter from KPNO filter set A was used for order separation. The detector was a Tektronix 2048 CCD (T2KB), with 27 $\mu$m pixels and 4 e$^-$ read noise.  The 300 $\mu$m slit width projected to 1.8\arcsec\ and 0.6\arcsec\ pixel$^{-1}$ in the spatial direction.  The full spatial dimension available covered 240\arcmin, with some vignetting from the echelle optics apparent toward the spatial extremes. The slit was placed across the XA region centered at coordinates RA = 20:57:16.25, $\delta$ = +31:02:28.2 (J2000) oriented east-west bisecting the `V' of XA, as shown in Figure~3.  An exposure of 1800 s provided spectral coverage of \Ha\ and [\NII] $\lambda\lambda$6548,6583 at a scale of 0.189\AA\ pixel$^{-1}$ in the dispersion direction.  This produced a velocity resolution of $\sim$17\vel.

The data were reduced in a standard manner using tasks in the `longslit' package in IRAF.  Wavelength calibration used Thorium-Argon lamp spectra and optical distortions across the dispersion were rectified using sky lines.

The resulting spectrum is quite striking.  Clumpy, bifurcated velocity structures are seen in each of the three main lines in the spectrum.  Figure~6 shows the \Ha\ line and the summed [\NII] lines magnified by a factor of three in wavelength and converted to velocity space.  The separation of the two components varies from $\sim$50\vel\ in the center of the field (intersected by HUT positions 1 and 2 as well as to the west of HUT1).  The velocity ellipsoids narrows to less than the resolution at the extreme east and west ends of the sampled region.   These data are discussed in more detail in section 4. 


\section{Analysis}

\subsection{The UIT B5 Filter - What Does it Really Show?}

It is important to understand exactly what the UIT-B5 image represents.  As  discussed in Paper~I, this assessment is complicated by a number of factors, and can be investigated from both theoretical and observational (morphological) considerations. From the theoretical side, shock models indicate that the dominant ionic species in the B5 band is often \CIV, which falls near the peak of the filter's efficiency.  However, while \CIV\ is the strongest species, it is far from the only one.  There are significant contributions from \OIII]$\lambda$1665 and \HeII$\lambda$1640 and some weaker lines.  Furthermore, the B5 filter profile encompasses $\sim$15\%  
of the total hydrogen two-photon flux (cf. Figure~10 of Paper~I).  The two-photon emission can be greatly enhanced if the shock encounters neutral gas, and the \CIV\ emission can be drastically reduced if the shock speed is less than 100 \kms, if carbon is badly depleted onto dust, or if a large \CIV\ optical depth scatters \CIV\ photons out of the line-of-sight.  The \CIV\ emission relative to the two-photon continuum is enhanced in the so-called incomplete shocks, where the shock interaction is (effectively) so young that the full cooling and recombination zone of a local shock front is not present.  As can be seen in Figure~1, the regions that show primarily as green and/or blue (and hence show little \Ha) are largely in this category.  

Multiplying IUE spectral line strengths by the filter throughput at those wavelengths, we have constructed a composite `B5 emission line' in various types of filaments as a point of reference.  This  composite spectral line should be what is sampled on average in the FUV image at regions where two-photon continuum is weak or absent.  We find  that for the brightest  apertures, \CIV\ and \OIII] each contribute roughly one third of the line flux seen in the B5 image with other lines making up the remainder.  In the lower surface brightness regions, the proportion of \CIV\ grows past 50\% 
while \OIII] correspondingly decreases toward 25\%.  
This provides evidence that  \CIV\ flux is more heavily attenuated by resonance scattering in  the brightest regions while intercombination lines benefit from increased column densities.  Cornett, et al. (1992) reached similar conclusions.


\subsection{Range of Shock Velocities}

The issue of shock velocities is important to the analysis of the XA region.  We have used the ``E'' shock models of Hartigan, Raymond \& Hartmann (1987, henceforth HRH) to assess shock velocity and possible variations.  The HRH models were calculated for planar shocks with smooth flow and complete recombination, with Herbig-Haro objects in mind.  They were calculated for a preshock density of $\rm n=100~cm^{-3}$ while  densities in the Cygnus Loop are probably an order of magnitude lower.  However, collisional deexcitation only begins to affect the UV emission around densities of $\rm n=10^{9}cm^{-3}$, so the HRH models are still reasonable for our analysis.

The high-ionization lines in our spectral range (\OVI, \OV, \NV, \CIV) are strong resonance lines.  Thus line ratio diagnostics involving any of them are not suitable for determining shock velocity without correction for resonance scattering, and these corrections can be highly uncertain.  Yet even the presence of these lines in our data indicates that shocks of at least a certain velocity must must exist in XA, since each of these ions only `turn on' as shocks of sufficient velocity ionize the gas to the necessary ionization state.  For instance, \OVI\ starts appearing in shocks near 160 \kms, rises rapidly, and stays strong even well above 200 \kms.  Hence the presence of strong \OVI\ emission indicates that shocks of at least 170 \kms\ must be present.  The \OVI\ lines are strong at all five HUT positions across the XA cloud but were not sampled by the IUE data.

While the existence of high ions suggests fast shocks, there are several indications of considerably slower shocks.  HRH models were fit to the six intercombination lines present in each of the IUE SWP apertures.  In most of these, we see that the best fit occurs at 120--140 \kms.  

\CII]/\CIII] is a good shock diagnostic for slow shocks.  Two of the four LWP spectra show no evidence of \CII] $\lambda$2325.  In the other two, we see a dereddened \CII]/\CIII] ratio of 0.4 (IUE2M) and 0.1 (IUE5S), a bit lower than the minimum ratios predicted by HRH of 0.5.  While the LWP camera was particularly insensitive to this part of the spectrum, it seems likely that much of the shocked material has not cooled enough to emit enough \CII] yet.  Nevertheless, the lowest model ratios occur between 110 and 140 \kms.

The ratio of \OIV] to \OIII] is another good shock speed diagnostic.  The ratio is insensitive to resonant scattering and elemental abundances, as both lines arise from the same element and both are intercombination lines.  Models predict the ratio rising from zero at 100 \kms\ and reaching a maximum value of 1.7 at 160 \kms.  However, the \OIV]$\lambda$1400 feature is blended with \SiIV\ and is unresolved by IUE.  HUT can resolve the two species, but without spatial information.  We are limited to saying that the ratio of \OIV] to \SiIV\ in this region is typically 3:1.  Correcting the IUE data by this factor, we get observed \OIV]/\OIII]=0.75--1.1, corresponding to velocities in the 120--140 \kms\ range.

Thus the \OVI\ emission implies a shocks at 160--200 \kms; the shock model fits, carbon ion ratios and oxygen ion ratios suggest slow shocks at 120--140 \kms; and the X-ray emission requires 300 \kms\ shocks.  Shocks slower than 120 \kms\ may well be present, but they produce little emission in our spectral range.  Given the morphological complexity of the region, it seems quite reasonable that a range of shock speeds must be present within our apertures.  

There are other possible explanations for the mixed signals we are receiving.  \OVI\ is present even in $10^6$ K gas, but given the emission measures derived by Miyata (1996) or Szentgyorgyi et al. (2000), the X-ray emitting gas would only account for $\sim$5\% 
of the \OVI\ observed.  Time-dependent ionization might increase this fraction somewhat, but the consideration of resonance scattering (\S3.3) implies that the intrinsic \OVI\ emission is a least twice as strong as a naive interpretation of the HUT spectrum indicates.  Another possibility is thermal evaporation (e.g. Teske \& Kirshner 1985), but Teske (1990) and Szentgyorgyi et al. (2000) have shown that evaporation is too slow to account for the relative positions of [\NeV], [\ion{Fe}{10}] and X-ray emission.  

A third possibility is a turbulent mixing layer (e.g. \cite{Slavin93}).  The surface brightnesses predicted by Slavin et al. fall two orders of magnitude short of those observed in XA when scaled to the pressure of the Cygnus Loop.  If XA is elongated along our line-of-sight (as implied by the resonance scattering analysis in \S3.3), the mixing layer emission could account for perhaps 1/10 of the observed fluxes.  A more general difficulty is that the Slavin et al. models predict \OVI\ fluxes at most $\sim$1.5 times (and more often comparable to) the fluxes of  of \HeII\ $\lambda$1640, \OIII] $\lambda$1664 and \OIV] $\lambda$1400, while the observed \OVI\ is an order of magnitude brighter when corrected for reddening and resonance scattering.

It is therefore likely we are dealing with a range of shock velocities even within each spectrograph aperture.  The `cloud' being struck by the blast wave at XA may be surrounded by a range of density instead of a hard boundary.  As the shock is driven into and around the cloud, a range of shock velocities would naturally result.  Alternatively, turbulence could result in shocks over a range of velocities and obliquities.  The optical images clearly indicate structure on scales smaller than the aperture sizes, again indicating that the apertures are likely sampling diverse conditions. 


\subsection{Resonance Scattering}

We now turn to a subject that has been known to affect UV line analyses for some time, but which is often difficult to quantify.  Optical depth effects caused by resonance line scattering--the absorption and re-emission of photons in strong resonce lines, affects our view of XA significantly.  Resonance line photons produced in a sheet of hot gas behind a shock tend to scatter out in the direction of lowest optical depth, i.e. perpendicular to the sheet.  To the extent that bright filaments in SNRs are sheets of gas seen edge-on (Hester 1987), the intensities of resonance lines will be suppressed relative to the intensities of forbidden or intercombination lines.  The intensities of lines from ions abundant in the ISM will also be attenuated between the SNR and observer.  Of the lines in the HUT and IUE spectra, the \OVI\ doublet, \NV $\lambda 1240$, \CII $\lambda$1335, \CIII $\lambda$977, \CIV $\lambda$1550 and \SiIV $\lambda$1400 will be most strongly affected.  An understanding of resonance scattering can be had by examining the ratio of resonance to intercombination lines of various species.  There are several ways we may implement this technique.

By comparing \CIII\ $\lambda$977 (from HUT) and \CIII]$\lambda$1909 (from IUE), we can see that considerable resonance scattering is present within all five HUT apertures.  Since both lines arise in the same ion, one needn't worry about abundance or ionization effects.  Furthermore, one line is a strong resonance line while the other is an intercombination line so the ratio of oscillator strengths -- $f_{977}/f_{1909}=4\times10^6$ -- is very high.  Figure~7 shows a model and observational comparison for this ratio.  Shock models show that the expected ratio of resonant to intercombination \CIII\ flux should be between 2.4 and 2.9 for the shock velocities seen in XA.  If we synthesize a \CIII] $\lambda$1909 HUT surface brightness by summing the IUE surface brightness for each HUT position and compare them to the dereddened \CIII\ $\lambda$977 fluxes, we see ratios between 0.6 and 1.4.  Clearly, there is considerable optical depth in \CIII\ $\lambda$977.  It could be interstellar absorption toward the Cygnus Loop, or it could be the optical depth of \CIII\ in the emitting gas along the line of sight.

To obtain better spatial resolution, we must resort to a ratio between two species of different ionization within the IUE spectral range alone.  \CIV\ and \CIII] have ionization energies  of 47.9 and 24.4eV, respectively.  Of the three high-ionization Li-like ions, \CIV\ has the largest oscillator strength and should be the most affected by resonance scattering effects while \CIII] has an oscillator strength six orders of magnitude smaller and should show negligible attenuation at any reasonable column depth.  Both are strong, uncontaminated lines resolved in even the IUE data.  In Figure~8, we again show the model/observational comparison similar to Figure~7 with the addition of north-south spectral profiles and [\OIII] image crosscuts at each position.  For shock speeds of 140 \kms\ and higher, we expect to see \CIV/\CIII]$\sim$3--4, yet at nearly all IUE apertures in XA, the observed ratio is between 0.6 and 1.4.  Never does it approach the value of 3 to 4 that we would expect for a shock which is producing high ionization emission.  This indicates clearly that very significant resonance scattering of \CIV\ is taking place, and not just in the brightest regions of optical emission, but also in more diffuse regions (the regions off the cloud core, toward the top and/or bottom of each HUT aperture).  In Figure~8, there is some evidence that the resonance scattering is more significant in the brightest  optical [\OIII] region, although the errors on individual points makes this difficult to quantify. 

Thus, compared to \CIII] $\lambda$1909 there are a flux deficits of approximately 2.5 and 3 respectively in \CIII\ $\lambda$977 and \CIV.  One optical depth of each would suffice to lower the flux by the factor observed.  Since we would expect more of the lower ionization \CIII\ to be present in the ISM than \CIV, and they appear to suffer equal extinction, we can conclude that the bulk of the optical depth is occurring within the line of sight to XA itself.  However, while we expect to see significant optical depth only in the bright filaments, we are seeing it everywhere! In the filaments, we see \CIV/\CIII]$<1$ but even in the interfilamentary  apertures (SWP~26481, S5M, S4S, S2S) we see \CIV/\CIII]$\leq2$.  This is probably because the optical depths in the \CIII\ and \CIV\ lines are $\sim$1 in the flow direction according to shock models (e.g. Raymond, Wallerstein \& Balick 1991), and even the apparently diffuse emission arises from shocks viewed only 10 or 20 degrees from edge-on.  Thus resonance scattering is still substantial, though less severe than in the sharper filaments.  In either case, one sees to about 1 optical depth in the resonance lines.  

Comparing again to the B5 image, one sees a good correlation between B5 emission and [\OIII] emission.  To the extent that the B5 image is dominated by \CIV$\lambda$1550, we can make a crude map of the \CIV/[\OIII] ratio.  The two ions have similar ionization energies and thus relative differences should be a valid resonance scattering probe.  This is discussed in Cornett et al. (1992) and Paper~I and is illustrated in Paper~I, Figure~9c.  The resonance scattering appears highest at the center of the cloud intersected by HUT position 2 and in the pair of filaments marking the eastern boundary of XA.  However, there is less contrast between the brightest and faintest B5 regions than there is in optical [\OIII].  This is consistent with \CIV\ being more affected by resonance scattering in the brightest (densest) lines of sight.  It is completely consistent with the lower resolution but more quantitative ratios in \CIII/\CIII] and \CIV/\CIII] discussed above.  If we see a factor of 3 in a ratio having good spectral resolution but poor spatial resolution (\CIV/\CIII] and \CIII/\CIII]) and a similar trend in a diagnostic having good spatial resolution but little spectral resolution (B5/[\OIII]), it seems highly likely that we would see a factor of 10 if we had good spatial and spectral resolution in the same observations.

Another estimate of the effects of resonance line scattering is provided by the \OVI\ doublet (e.g. Long et al. 1992).  The intrinsic intensity ratio of the 1032\AA\ and 1038\AA\ lines is 2:1, and so is the ratio of opacities.  If photons are scattered out of an emitting sheet of gas in a regime of modest optical depth, the observed ratio will be less than 2:1. The ratios of the intensities in Table~2 range from 1.0 to 1.3, with only the HUT5 spectrum allowing a 2:1 ratio within errors.  To check the severity of the problem, we can compare the \OVI\ intensities to the intensity of the [\NeV] $\lambda$3425 line from Szentgyorgyi et al. (2000) by adding up the fluxes in the [\NeV] image in the patches corresponding to the HUT apertures.  The observed \OVI/[\NeV] ratios range from 10 to 30, or 20 to 60 after correction for reddening.  The HRH shock models predict ratios of above 100 for the velocity range where both lines are strong (P. Hartigan, private communication), suggesting \OVI\ attenuation by roughly an order of magnitude.  Within the HUT spectra themselves, the detection of \NeV] $\lambda$1146 and the upper limit on \ion{Ne}{6}] $\lambda$1006 are at about 1/50 the brightness of the \OVI\ doublet.  The HRH models predict these lines 30 to 150 times fainter than \OVI.  This implies \OVI\ attenuation no greater than a factor of 3, suggesting an optical depth of order 1.

As a consistency check, we note that the observed \OVI\ surface brightness, corrected for reddening but not for resonance scattering, requires $\rm n_e N_{O VI}~\sim~5 \times 10^{15}~\rm cm^{-5}$ for the \OVI\ excitation rates at temperatures of $10^{5.4}$--$10^6$K.  Densities in the XA region were derived by Szentgyorgyi et al. (2000) based on ROSAT and [Ne V] images. Scaling that density to the revised Cygnus Loop distance of 440 pc, and assuming pressure equilibrium between the X-ray emitting gas and the cooler \OVI\ emitting gas, we estimate $n_e~\sim~6~\rm cm^{-3}$, so $\rm N_{O VI} $ is a few times $10^{14}~\rm cm^{-2}$.  From the Figure~6, we can use a line width of 50 \kms\ to account for bulk motions in this turbulent region, the optical depths of the 1032 and 1038\AA\ lines are about 2 and 1, respectively.

Thus we confirm that the estimate of $\sim$ 1/3 for \OVI\ attenuation by resonance scattering is consistent with the densities estimated from X-ray and optical analyses.  Interstellar absorption of the \OVI\ doublet is expected to exist, but to be considerably weaker than is the case for \CIII\ $\lambda$977.

\subsection{UV emission from the X-ray emitting gas}

The [\ion{Si}{8}] lines at $\lambda\lambda$1441, 1445 have not been previously reported in supernova remnant spectra, though they are observed in the solar corona (e.g. Feldman et al. 1997).  In ionization equilibrium, the emissivities of these lines peak at $8\times10^5$K.  The ROSAT X-ray spectra of the XA region show unusually cool temperatures for the Cygnus Loop, so the [\ion{Si}{8}] emission might be expected to detectable.

The combined emissivity of the the \ion{Si}{8} lines peaks at $2.3\times10^{-26}~\rm erg~cm^6~s^{-1}$ according to CHIANTI (Landi et al. 1999) assuming the Mazzotta et al. (1998) ionization equilibrium and Allen (1973) silicon abundance, with excitation rates from Bhatia \& Mason (1980).  Departure from ionization equilibrium cannot greatly enhance this emissivity, because the Mazzotta et al. calculations show a fairly large fraction of silicon in this ion near its peak abundance.

From the ROSAT PSPC spectra of 1 arcminute square regions encompassing the HUT aperture positions, Szentgyorgyi et al. (2000) estimated temperatures of $1.0-1.2 \times 10^6$ K and emission measures of 40 $\rm cm^{-6}~pc$.  Their single temperature, ionization equilibrium models are open to question, of course.  Miyata (1996) analyzed ROSAT and ASCA spectra of a 5 arcminute diameter patch including XA.  Her two-temperature models showed temperatures of 0.043 and 0.267 keV, with emission measures of 17,000 and 36 $\rm cm^{-6}~pc$, respectively.  However, the enormous amount of relatively cool gas would imply far brighter [\NeV] and \OVI\ emission than observed.  

Taking 40 $\rm cm^{-6}~pc$ as typical of the X-ray emitting gas, we would expect about $7 \times 10^{-17}~\fluxden$, or about half the observed average flux after correction for reddening.  Thus it appears that a substantial emission measure of gas near or just below $10^6$ K is present and that silicon is not severely depleted in the X-ray emitting gas.  Estimates of depletion are complicated by the fact that Si and Fe dominate the ROSAT 0.25 keV band, so that severe depletion would imply an increase in emission measure to account for the observed X-ray flux.  Nonetheless, it would be difficult to reconcile a depletion by more than a factor of 2 or 3 with the observed [\ion{Si}{8}] flux.  According to the models of Vancura et al. (1994) for dust destruction in a somewhat faster shock, half the silicon is liberated from grains at a column depth of about $4\times10^{18}~\rm cm^{-2}$, which corresponds to an elapsed time of 4000 years with the density of 3.4 estimated by Szentgyorgyi et al. (2000).  This is in sharp contradiction with the estimated interaction age of $\sim$500 years (\S4).  Silicon should not have been liberated in such quantities yet and suggests grain sputtering may be substantially faster than predicted.  Though the [\ion{Si}{8}] feature is marginally detectable in the individual HUT spectra, there does not appear to be a gradient in its intensity moving west from the shock.  Decreasing X-ray temperature may be balancing any increasing silcon abundance.

In summary, the \ion{Si}{8} emission is probably formed in gas just below $10^6$K having an emission measure somewhat above that of the X-ray emitting gas a slightly higher temperature.  The silicon in this gas must have been mostly liberated from grains.  As mentioned above, the X-ray emitting gas accounts for only a few percent of the \OVI\ emission seen by HUT.  The cooler \ion{Si}{8} emitting gas could account for $\sim$20\%.


\section{A New Picture of XA} 

It is now widely held that the Cygnus Loop was created by a cavity explosion of a B0-B1 star (Levenson et al. 1998, and references therein).  Through ionizing radiation and mechanical wind power, the precursor excavated a roughly spherical cavity some 10 pc in radius.  \cite{Levenson98} proposed an overall pre-SN structure for the ISM surrounding the Cygnus Loop composed of an overall diffuse atomic shell interrupted in several places around the surface by denser clouds.  One of these clouds is responsible for the northeast radiative filaments (NGC~6992) and the XA region (NGC~6995) where it projects into the interior of the cavity.  We adopt this model here and propose additional details which could result in the morphology we observe at XA.

The entire surface of the blast wave as marked by the X-ray edge is indented at several points along the SNR perimeter.  This is evidence that the blast wave is interacting with clouds large enough to cause a significant impediment to the shock front, not just a ``dimple''.   The indentation in the southeast is one such example; the blast wave has just begun to interact with the tip of a large cloud.  Balmer-dominated filaments can be seen diffracting around the unseen cloud and a bright knot of UV/optical emission is seen as the Southeast Cloud (\cite{Fesen92}, \cite{Graham95}).

The XA region is another such indentation (Hester \& Cox 1986), though likely a bit more complicated in geometry  and advanced in effective `age'.  The entire surface of the blast wave is indented like a soccer ball being kicked.  According to Levenson (private comm.), such a sharp, indented X-ray edge would require a cloud of material several parsecs long in our line of sight, and this agrees with the 3 pc depth estimate of Szentgyorgyi et al. (2000).  This ``kicked soccer ball'' model explains why we see such an X-ray peak in the area--as the blast wave encounters the dense material at the pre-SN cavity wall, the increased density and pressure enhance thermal X-ray emission.  The relatively cool temperature of the X-ray emitting gas in XA also suggests a density enhancement.  Since the blast wave is indented in a direction roughly in the plane of the sky, we have a long line of sight through the current rim of the SN.

We propose that there are many fingers of denser gas sticking out from the cloud responsible for the bulk indentation mentioned above (in keeping with the kicked soccer-ball analogy, perhaps these should be called `toes').  We submit that the XA `cloud' proposed by Hester \& Cox is in actuality one of these fingers of higher density cavity wall material protruding into a more diffuse region to the west as pictured schematically in Figure~9.  The origin of this finger is unknown, but it seems possible it is just a random density fluctuation in a large cloud surface.  The expanding wind from the precursor may have further shaped the cloud topography.  While the two situations are undoubtedly different, the famous ``elephant trunk'' structures in the HST image of the Eagle Nebula are of similar dimensions to the proposed XA finger.  

Detailed hydrodynamic simulations of shock-cloud interactions have been performed and many features present in the models are seen qualitatively in XA.  A set of 2-D, cylindrically symmetric numerical models were calculated by Bedogni \& Woodward (1990) with conditions similar to what we believe to exist in the XA region (v$_{ex}$=400 \kms, M=26.4, $\rho_{cloud}/\rho_{ambient}$=2-20, their cases 4 and 5).  The main features of these models are a cloud shock, a reverse shock and the overall diffraction of the main SN blast wave.  While these models employ spherical clouds, it is conceptually easy to extend these models to a non-spherical or finger geometries.

From simple inspection, XA appears to be in a stage similar to Bedogni \& Woodward's Figure~4a (though elements of both earlier and later stages may be present).  The cloud appears to be oblong, oriented in the east-west direction approximately 45\arcsec$\times$90\arcsec (0.1$\times$0.2pc).  We see a bright, `3'-shaped region of emission at the upstream (western) end of the cloud--presumably the transmitted cloud shock seen in the models.  To the east are a strong pair of curved shocks flanking the cloud closely resembling the diffracted blast wave of the models.  An assumed blast wave speed of 400 \kms\ gives an interaction age of roughly 500 years or roughly a tenth the age of the entire remnant.

Models predict that the blast wave will diffract around the cloud.  We see this as a  faint Balmer-dominated filament in the Southeast Cloud (\cite{Fesen92}; \cite{Graham95}; Paper~I).  In XA, however, we see a series of filaments strong in ionic lines ([\OIII], [\SII], \CIV, [\NeV]);  unmistakable signs of a radiatively cooling shock.  This could result from an extended atmosphere of relatively dense material away from the XA finger itself, or it could be due to the lower effective speed of an oblique shock.  Instead of a single set of diffracted filaments, we see a series of quasi-parallel filaments bent around the cloud.  This is most apparent on the northern side of the finger where a bright FUV/[\OIII] filament is seen followed a short distance later by a bright \Ha\ filament.

As the blast wave first encounters the inclined finger, it sets up a slow cloud shock at the tip.  Meanwhile the point of contact between cloud and blast wave races down the slope of the inclined finger.  At the present time, the blast wave has covered $\sim$0.2 pc while the cloud shock, progressing more slowly through the denser material, has reached about HUT position 2.  We see the most recently shocked material in the east, emitting in \CIV\ and [\OIII], while the material which was shocked first has cooled enough to emit strong \Ha\ to the west.  At the extreme east is the bright north-south [\NeV] filament seen by Szentgyorgyi et al. (1999), which may represent the leading edge of the main blast wave that has wrapped around the cloud and is proceeding eastward.  As shock velocity is likely 150 \kms\ or greater, thermal instabilities are probably an issue as well.  Indeed, the cooler \Ha\ filaments appear more turbulent than the warm, smoother [\OIII] morphology.

The shape of the high-dispersion spectra (Figure~6) lends some support to this conclusion.  The brightest portion of the spectrum, which undoubtedly represents the slow cloud shock, appears at the eastern edge of HUT1 in both lines.  Both positive and negative velocity components appear bowed outward into a velocity ellipsoid.  Our interpretation of this is that the blastwave is driving oblique shocks into the denser finger.  We see blue-shifted emission from the back side of the cloud and red-shifted emission from the front.  If we take the presence of \OVI\ to indicate a shock speed of 170\kms, the postshock material will be moving at 3/4 that speed or 130\kms.  We see velocity components at $\pm$25\kms\ indicating that the front and back sides of the cloud must be at an angle of around $\rm\pm10^o$.  This could indicate that the cloud is very roughly conical with an opening angle of 20$\rm^o$.  

Since cooling rate is proportional to density, the dense post-shock material of the finger will cool much more quickly than the more diffuse material surrounding it.  This is manifested by an extended region of crushed, tangled  \Ha\ ``fluff'' seen to the west of the cloud shock.  This is the dense, post-shock gas cooling rapidly from the cloudshock.  In the echelle spectra, we see a clearly bifurcated velocity structure which coincides with the ``fluff'' seen in Figure~1.  It is interesting to note that in both spectral lines, the redshifted (positive velocity) component is brighter east of cloudshock while it is the blueshifted component which is brighter to the west.  We conjecture that this `cross over' implies brighter emission on one side of the cloud the other;  the east side (ahead of the cloud shock) is currently being contracted along our line while the west side (behind the cloud shock) is expanding and cooling. 

The hydrogen recombination time is $\rm0.2-1 \times 10^5 / n_e$ years for 1000--10,000 K gas.  We do not have a density measurement, but pressure equilibrium with the X-ray emitting gas would imply $\rm n_e~=~200 - 2000~cm^{-3}$  in this temperature range.  These densities would predict that recombination is largely complete within about 15" behind the shock, much less than observed.  However, magnetic pressure support could reduce the density of the cool gas (e.g. Raymond et al. 1988) and thereby increase the distance between the current location of the shock and the remnants of the recombining gas.  Models of the interaction of a blastwave with a cloud indicate that dynamical instabilities wrap up the magnetic field lines until magnetic forces halt the growth of the turbulent cascade (MacLow et al. 1994).  Thus we expect that the ``fluff'' to the west of the XA bright knot should show a wide range of densities resulting from different degrees of magnetic support, along with a velocity field indicative of the turbulence generated by the blastwave-cloud interaction.  To the east of the cloudshock, we see emission decrease steadily, probably indicative of less complete cooling.  The velocity structure appears narrower as well. 

Without much more detailed spectra and imaging, it is unwise to make more specific arguments, but certain corrugations and irregularities on the cloud surface and patterns of density enhancement might produce the other observed features.

If the cloud `surface' is indeed conical, we should see the blast wave driving shocks into the finger along its entire surface.  While the bright cloud shock is still working its way down the length of the finger, there is a fainter ``tube'' of  emission stretching from the bright cloud shock to the east toward the X-ray edge.  These may be either secondary shocks seen limb-brightened against the north and south edge of the finger or bright filaments seen less than edge on across the face of the cloud.  The latter is consistent with the observed optical depth patterns seen in \CIII/\CIII] and \CIV/\CIII].  There would also be slower, secondary shocks being driven into the cloud by fluid instabilities along the faces of the finger.  This multitude of shocks combined with the large size of our spectrograph apertures then explains why we see a range of shock velocities.

The final feature found in the numerical simulations but absent from our new picture of XA is that of a bow-shock.  The simulations predict that a prominent reverse shock should be set up along with the cloud shock;  this is nowhere in evidence.  The Mach number for the postshock gas is certainly high enough to form a stationary bow shock by the time the main blast wave reaches the back of the cloud  (Spitzer 1982, Bedogni \& Woodward 1992).  No bow shock is seen in any of the optical or UV images and the X-ray observations show no intensity concentrations in the area--quite the opposite, in fact.  However, all the simulations were done using sharply bounded clouds with substantial density contrast.  Our model involves a much softer density gradient, which may disrupt the formation of a detectable bow shock.

There are two more bright patches of emission seen in Figure~1 to the north of XA and at least one to the south.  The long line of tangency through the blast wave brings with it considerable line of sight confusion.  It is clear that several of the features seen in the XA field must be unrelated physically.  For instance, the long nonradiative filament curving from north to south through the middle of Figure~1 is likely a ripple in the front or back side  of the blast wave and not directly related to the bright radiative emission in XA.  Similarly many areas of ionic emission are likely radiatively cooling shocks seen obliquely.  The other nearby `cloudlets' are presumably similar fingers like XA at different places along the line of sight, different physical scales and in different stages of evolution.  The bright spot immediately to the north-east of of XA could well be a similar finger in the very early stages of shock interaction.

\section{Summary}

The Cygnus Loop is a cavity explosion 5-10,000 years old.  For most of its life the SN shock wave has been propagating through this rarified cavity.  Only recently has it encountered the denser cavity walls and begun to decelerate.  In regions where the cavity wall is atomic and relatively diffuse, deceleration is slow and we see smooth, nonradiative, `Balmer-dominated' filaments.  In regions where denser clouds intrude into the walls of the cavity, deceleration is stronger and the shock emits strongly in X-ray, optical and UV lines.  

The strong optical and FUV line emission, as well as the bright, steep X-ray gradient shows that the XA region is produced by a shock collision with a dense cloud of some sort.  The fact that the XA region is significantly indented from the circularity of the rest of the remnant implies that the shock has encountered a cloud large enough to decelerate it along a large portion of the shock surface.

Using a wealth of FUV spectral and imaging data we have investigated the nature of XA in some detail.  Imagery in \Ha, optical [\OIII], a FUV band centered at 1500\AA, and soft X-rays gave us morphology information at a wide range of temperatures.  A combination of moderate resolution HUT spectra and grid of spatially resolved, low resolution IUE spectra allowed us to measure fluxes from various ions in the FUV including \OVI, \NV, \CIII, \CIII], \OIV, \OIII], and others as they varied across the field.  Finally, a high-resolution echellogram in \Ha\ and [\NII] gave us velocity information in a slice through the field.  From these data we have drawn the following conclusions:

1 -- In comparison to radiative shock models, certain FUV line ratios suggest shock velocities of 140 \kms.  However, the existence of high-ionization species requires a shock velocity of 170 \kms\ or higher.  We resolve this by noting that the shock structure in this region is extremely complicated.  Our spectrograph apertures are large enough that we are undoubtedly sampling both fast and slower shocks at the same time, thus leading to the confusion.   

2 -- The filaments we see are shock fronts seen tangent or nearly tangent to our line of sight.  Thus we expect to see the highest optical depth in these filaments where there is a long path length in which to absorb radiation.  Ratios between resonance and intercombination lines (\CIV/\CIII], \CIII$\lambda$977/\CIII]$\lambda$1909, etc.) show that this is true.  But the optical depth in certain FUV lines is on order unity even far from the filament cores.

3 -- For the first time in a SNR, we see certain very weak lines including [\ion{Si}{8}] emission at 1443\AA.  This ion likely arises in nearly million-degree gas and is likely material liberated from grains.  Liberation of Si from grains appears to be much faster than expected from the cononical sputtering rates.

4 -- We have proposed that the triangular XA knot itself is not an isolated cloud as previously thought, but a projecting finger of material sticking out of the dense cloud which causes the indentation in the northeast perimeter of the Cygnus Loop.  While the appearance of the filaments suggests a cloud with a diffracted shock wave, all related filaments seen in the field are radiative in nature suggesting an extended atmosphere of relatively dense material away from the finger itself.  The high and widely distributed resonance scattering supports this conclusion as well.  Other regions in the field can be interpreted as similar shocked fingers of different sizes at different places along the line of sight.  These knots represent earlier or later stages of evolution than XA, depending on the time since encountering the shock.


\section{Acknowledgements}  
The authors would like to acknowledge the discussions with Ravi Sankrit, David Strickland,  Nancy Levenson, Sally Oey, Kenneth Sembach and Robin Shelton.  Susan Durham provided the \Ha\ filter.  Special thanks to the HUT and UIT teams and to the astronauts of STS-67 (Astro-2) for their diligent work to obtain the HUT and UIT data presented herein.  This research was supported by the Center for Astrophysical Sciences at Johns Hopkins, NASA contract NAS5-27000, and NAG8-1074 to the Smithsonian Institution.



\newpage

\begin{figure} 
\epsscale{.75}\plotone{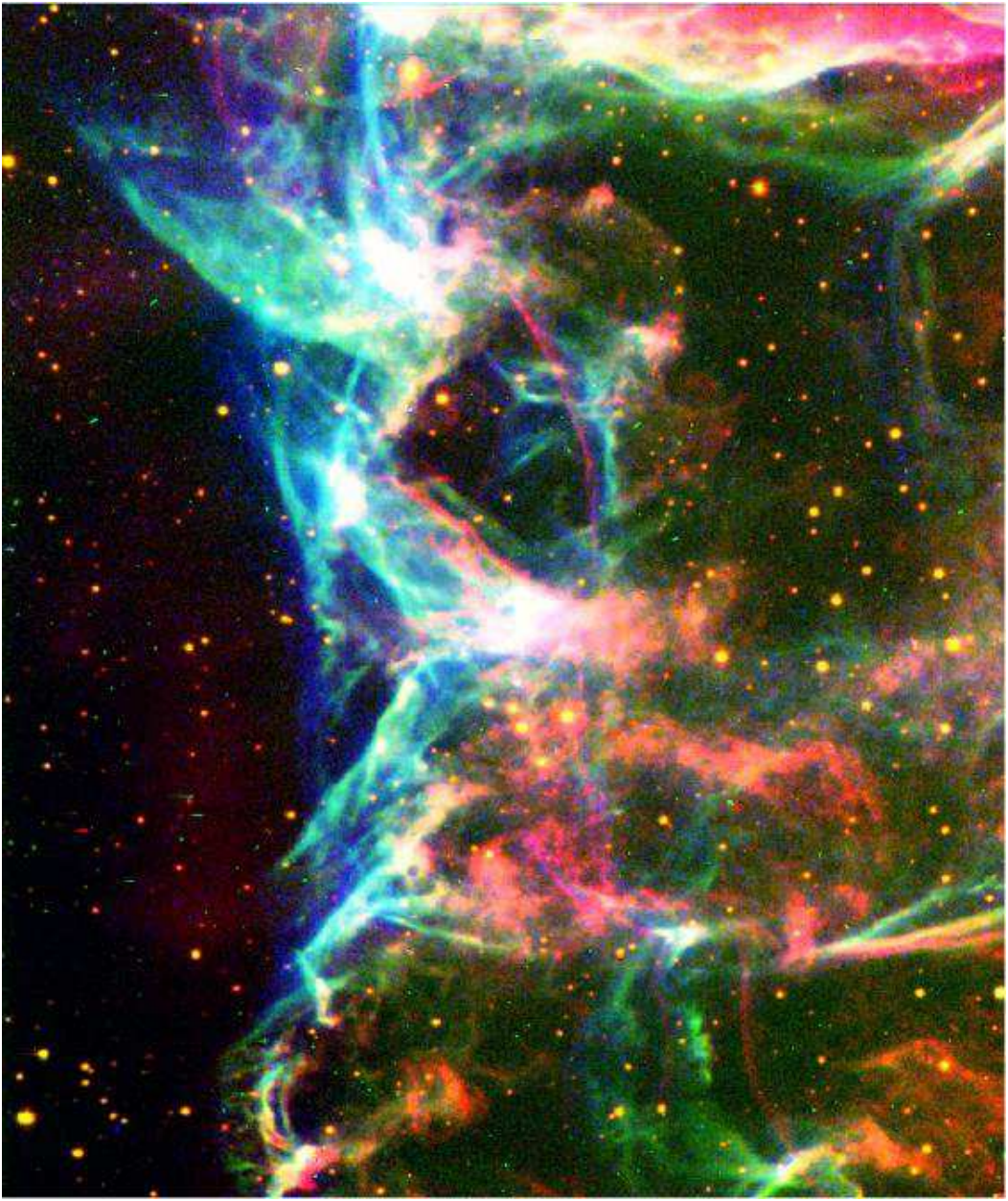}
\caption{A three-color image of XA. H$\alpha$ is represented as red,
[\ion{O}{3}] as green, and the FUV B5 band is blue.  All images are displayed
logarithmically and color levels have been adjusted arbitrarily to avoid
saturation.  The XA cloud core is the bright region in the center of the field.
The tip of the `Spur' filament studied by Raymond et al. (1988) is visible at
extreme upper right.  The incomplete shock filament studied by Blair et
al. (1991b) is the upper bright patch.  North is at the top, east to the left
and the image is 10\arcmin\ across.}
\end{figure}

\begin{figure} 
\epsscale{.5}\plotone{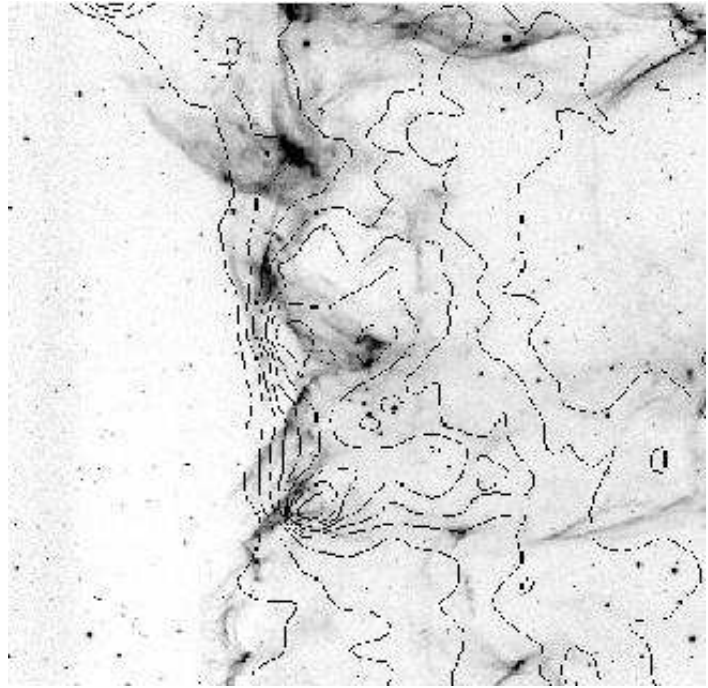}
\caption{ROSAT HRI contours overlaid on our [\ion{O}{3}] image.  Contour
interval is 0.1 of the X-ray maximum.  Notice the sharp drop-off to the east
and the two bright regions roughly concurrent with bright [\ion{O}{3}]
filaments.}
\end{figure}

\begin{figure} 
\epsscale{1}\plotone{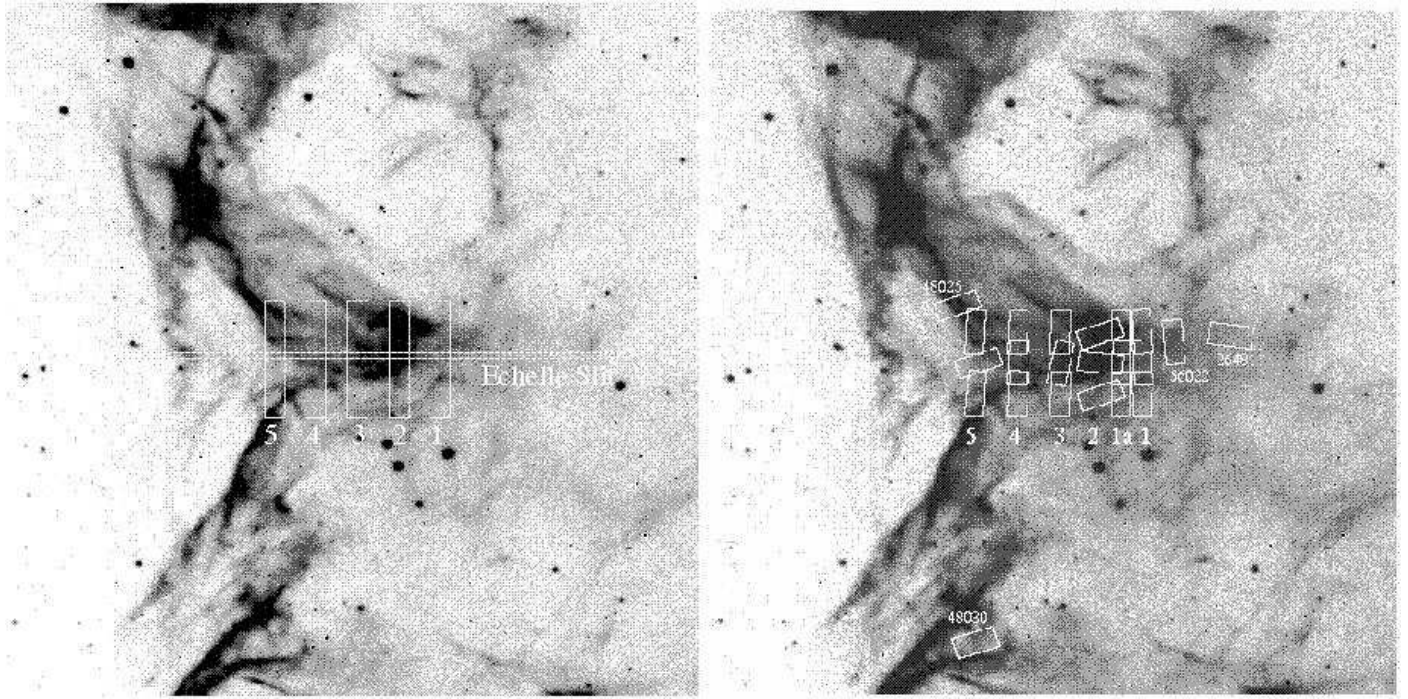}
\caption{Positions of the (a) five HUT apertures (10\arcsec$\times$56\arcsec,
left) and the echelle slit and (b) the 22 IUE apertures
(10\arcsec$\times$20\arcsec, right) both overlaid on our [\ion{O}{3}] image.
North is at the top and east at the left.  See Tables~1 and 3 and the text for
details.}
\end{figure}

\begin{figure} 
\epsscale{1}\plotone{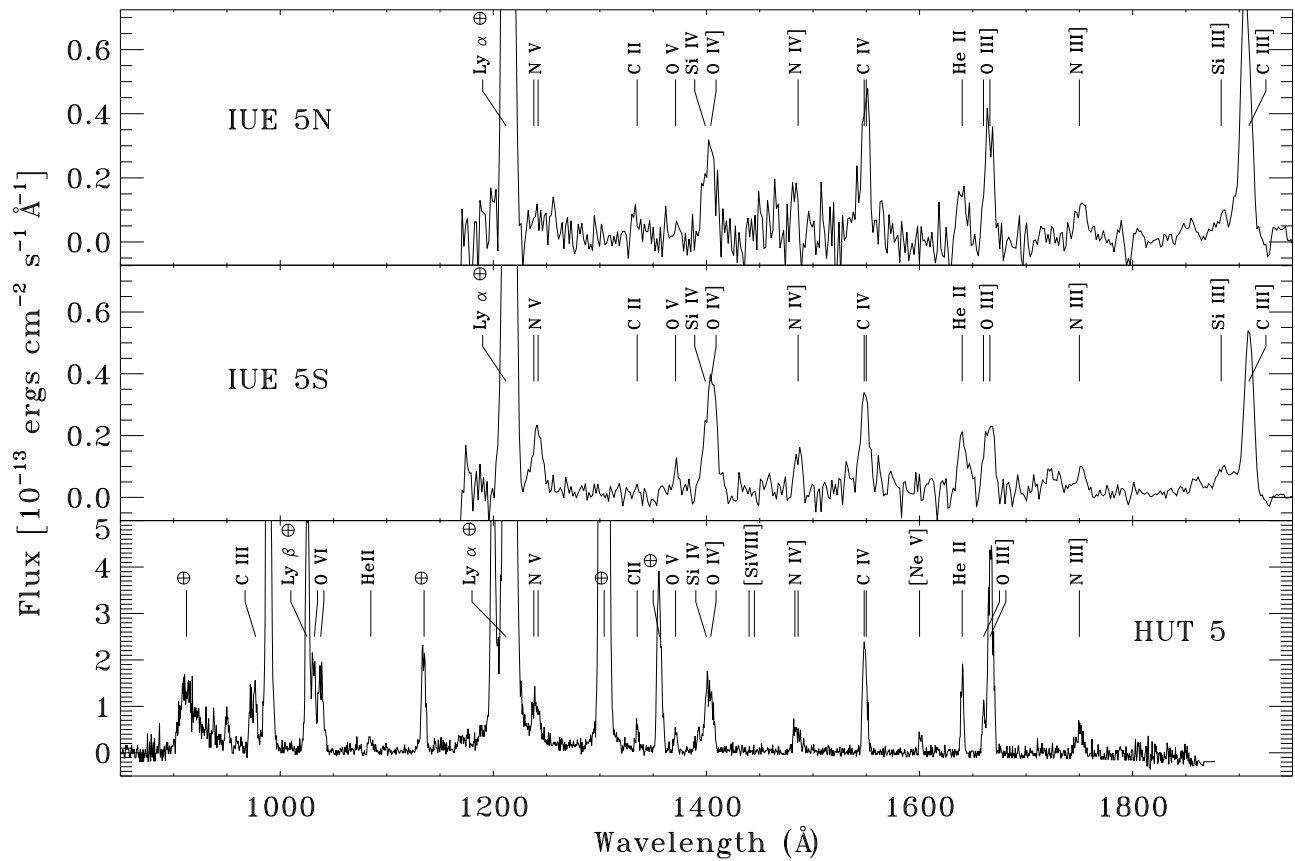}
\caption{A comparison of HUT and IUE spectra at HUT position 5 with line
identifications.  The top panel shows IUE position 5N, a typical lower
ionization region straddling a bright optical filament.  The middle panel shows
IUE position 5S which is of higher ionization, though similar total intensity
to S5N.  The bottom panel shows the HUT spectrum for position 5 for
comparison.}
\end{figure}

\begin{figure}
\epsscale{1}\plotone{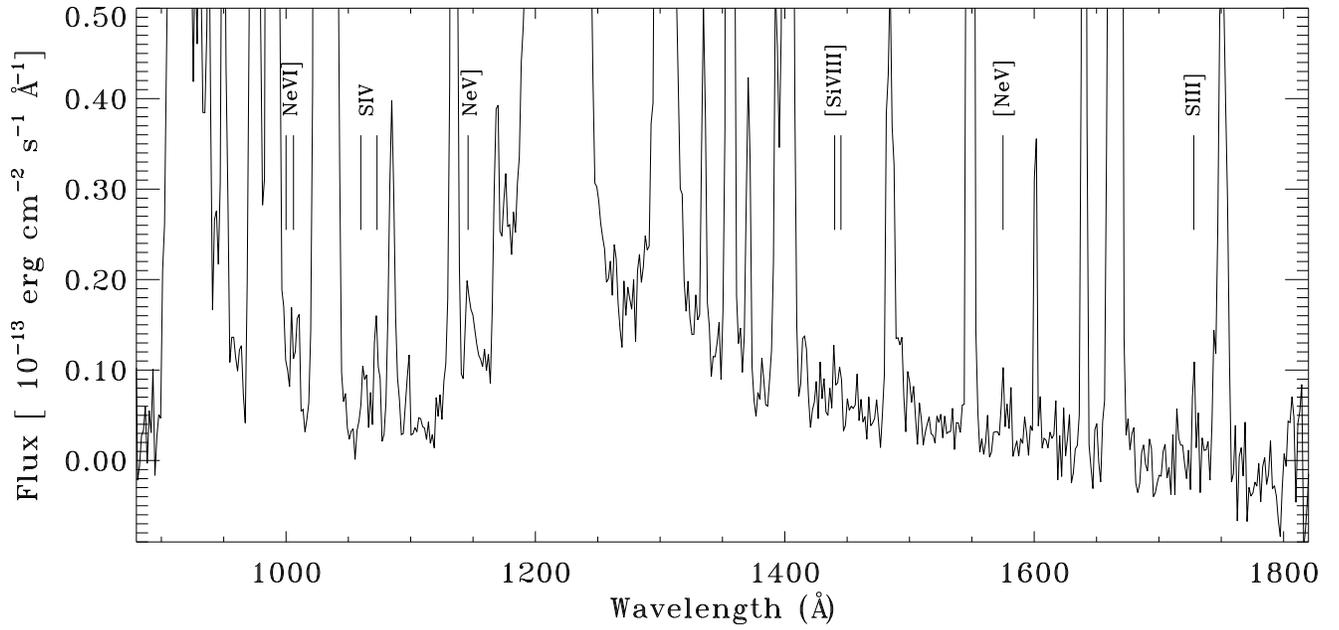}
\caption{By adding the five HUT exposures together we can create probably the
highest S/N spectrum of a SNR ever taken in the FUV with a total exposure time
of 2320 seconds.  Certain very weak spectral features pop out in this spectrum
which are otherwise lost in the noise including the never before seen
[\ion{Si}{8}]$\lambda\lambda$1440,1445.  Note, the \ion{Ne}{6}]
$\lambda\lambda$999.6,1006.1 lines are indicated here, but were not detected.}
\end{figure}

\begin{figure}
\epsscale{.75}\plotone{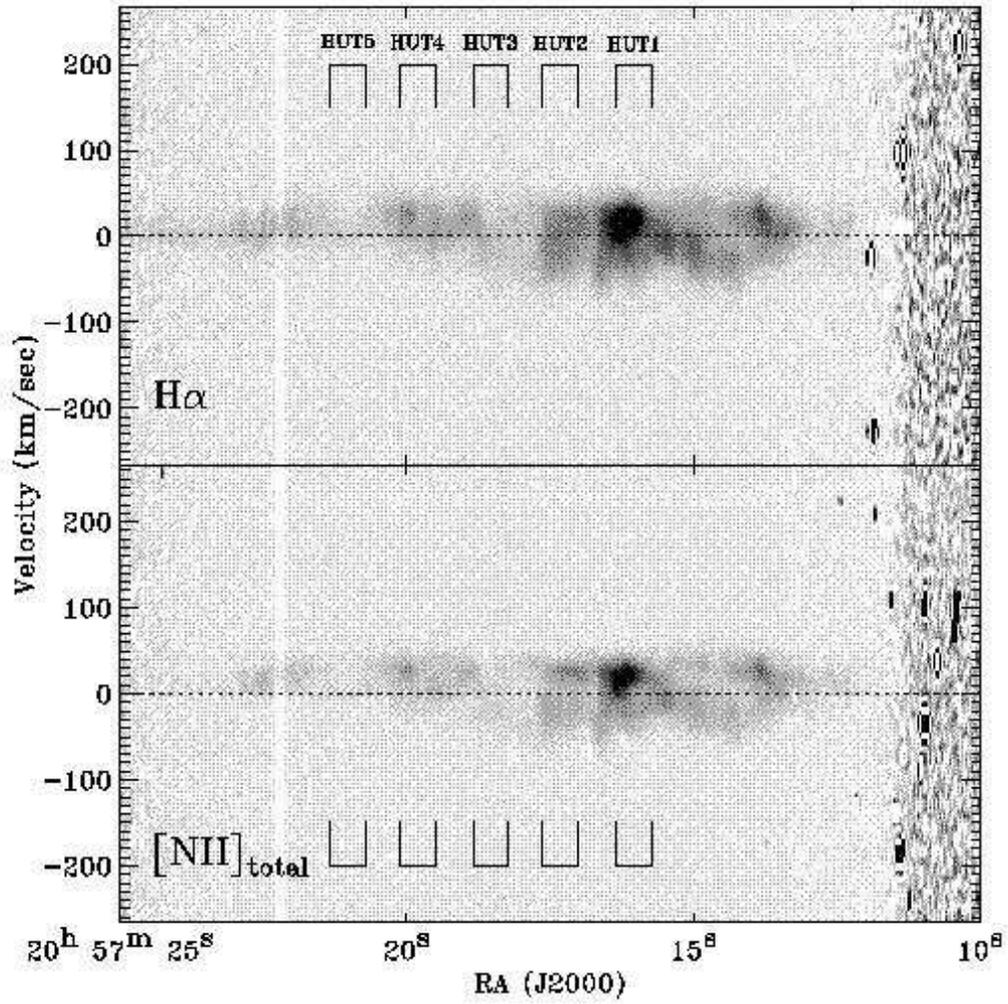}
\caption{Longslit echelle spectra in H$\alpha$ and combined [\ion{N}{2}]
$\lambda$6548+$\lambda$6583.  The positions of the HUT apertures are indicated.
The image has been magnified by a factor of three in the velocity direction.
Note the distinct bifurcated structure visible in both lines.  The maximum
separation of the centroids of the two components is $\sim$50 \kms.}
\end{figure}

\begin{figure} 
\epsscale{1}\plotone{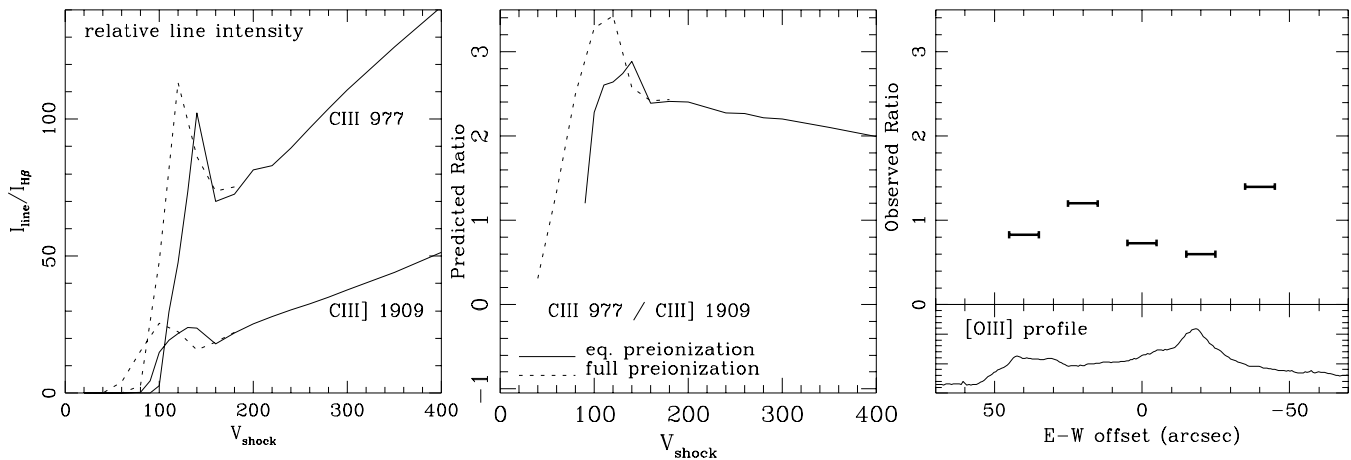}
\caption{Model predictions and observations of the ratio of \protect\ion{C}{3}
$\lambda$977/\protect\ion{C}{3}] $\lambda$1909. This ratio provides information
on resonance line scattering.  Unfortunately, our spatial resolution is
lessened by the necessity of using only full HUT apertures.  The left panel
shows line intensity as a function of velocity.  Line intensities are relative
to I(H$\beta$).  The middle panel shows the line ratio expected from shocks of
different velocities assuming equilibrium preionization (solid curve) and full
preionization (dashed curve).  The right hand panel shows the ratio of
dereddened line intensities, which are considerably lower than the model
predictions.  The bottom panel shows the optical [\protect\ion{O}{3}]
brightness profile along the same region covered by the HUT slits, and some
anti-correlation is seen of the ratio vs. [\protect\ion{O}{3}] intensity. This
indicates more resonance scattering in the brighter optical filaments.  Of
order one optical depth of resonance scattering will produce the observed
ratios.}
\end{figure}

\begin{figure} 

\epsscale{1}\plotone{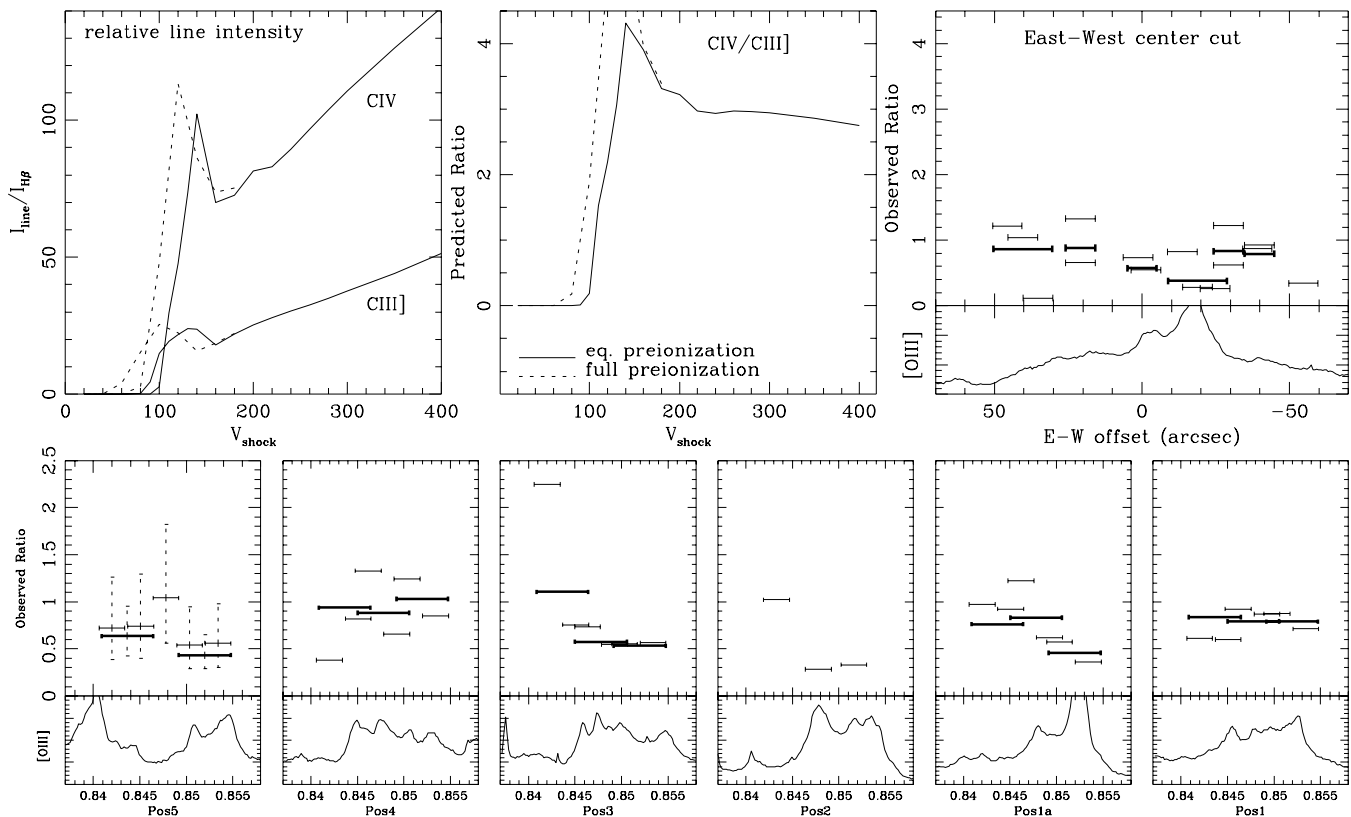}
\caption{Model predictions and observations of the ratio of \protect\ion{C}{4}
to \protect\ion{C}{3}], another resonance scattering diagnostic.  The top three
panels are in the same format as Figure~7.  Since both ions are available in
the IUE bandpass, greater spectral resolution is possible and we can show ratio
variations along each HUT position (the bottom six panels).  The heavy bars
represent full-aperture IUE data.  Lighter bars are for half-apertures (see
text).  North is at the left in each of these and the HUT5 position plot
(leftmost) shows typical ratio errors.  [\protect\ion{O}{3}] profiles are shown
below each ratio plot.  For any reasonable shock velocity, the ratio of
\protect\ion{C}{4} to \protect\ion{C}{3}] should be greater than 3.  However,
the observed ratios are closer to unity, implying extensive \protect\ion{C}{4}
resonance scattering in the HUT and IUE apertures.}
\end{figure}

\begin{figure} 
\epsscale{1}\plotone{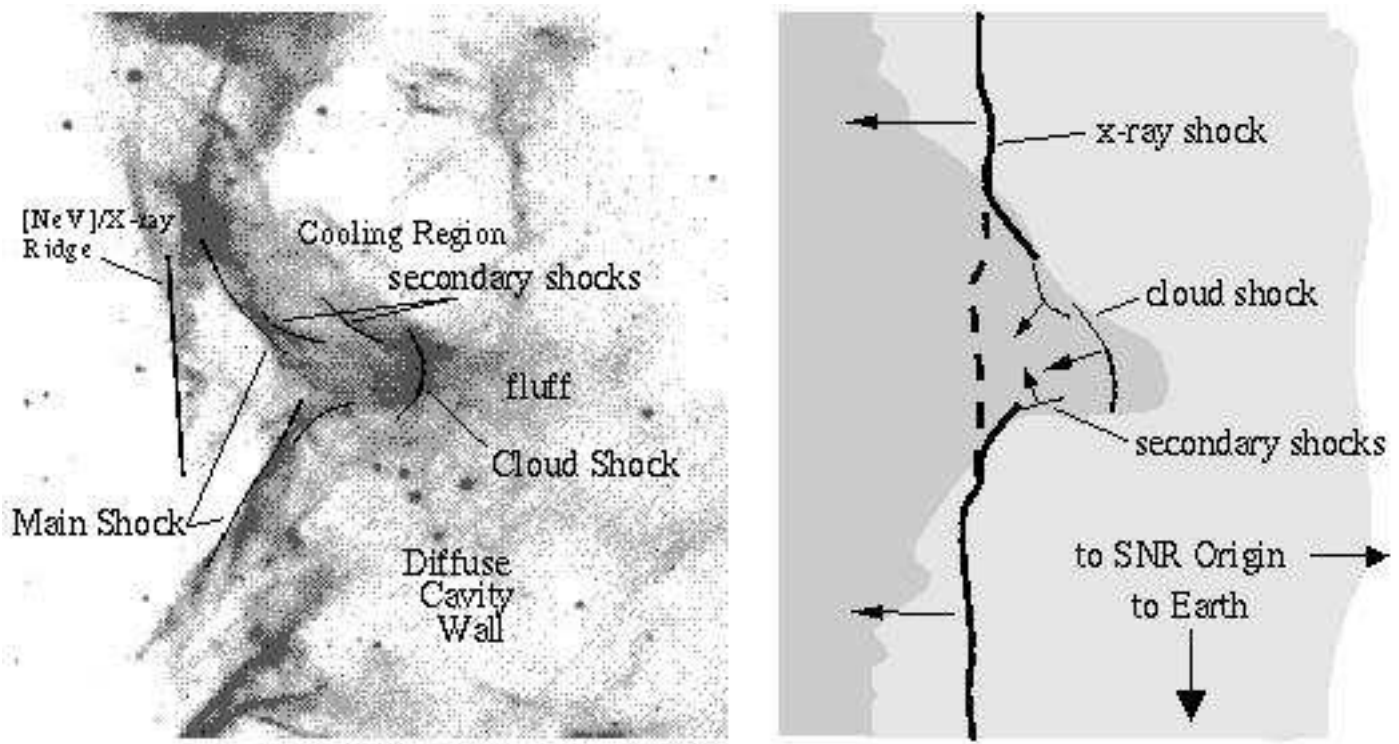}
\caption{Our proposed physical interpretation of the XA region is shown
schematically.  On the left we present an overlay on our [\protect\ion{O}{3}]
image defining the features referred to in the text.  On the right is
illustrated a view in the plane of the sky of our model.  The main blast wave
has recently started interacting with the denser medium of the cavity wall.  A
finger of this denser material protrudes into more diffuse surrounding cavity
material.  A cloud shock is set up in this denser finger while faster,
radiative shocks are set up in the diffuse cavity wall material.  Secondary
shocks propagate into the finger from the sides.  The `fluff' is composed of
lower-ionization post-shock material that was shocked longer in the past.
Meanwhile, the main blast wave continues to propagate through relatively
diffuse material.  This is what we see as the bright X-ray and
[\protect\ion{Ne}{5}] ridge.}
\end{figure}

\newpage

\begin{table} Table 1: HUT Observation Log\\
\begin{tabular}{lllll} 
ID  & RA (J2000)& Dec (J2000)& Obs. Date& Exposure\\
\tableline 
HUT1 & 20:57:14.72& +31:02:28& 10-Mar-95 & 348 s\\
HUT2 & 20:57:16.33& +31:02:28& 10-Mar-95 & 400 s\\
HUT3 & 20:57:17.84& +31:02:29& 10-Mar-95 & 504 s\\
HUT4 & 20:57:19.43& +31:02:29& 13-Mar-95 & 498 s\\
HUT5 & 20:57:20.94& +31:02:29& 13-Mar-95 & 570 s\\
\tableline 
\end{tabular} 
\end{table}

\begin{table} Table~2: HUT Line Fluxes and Uncertainties$\rm ^a$\\
\begin{tabular}{lrlllll} 
Line      &\AA & ~~~~HUT1     & ~~~~HUT2     & ~~~~HUT3     & ~~~~HUT4     & ~~~~HUT5     \\
\tableline 
\ion{C}{3}& 977& 1.552 (0.066)& 1.927 (0.177)& 1.232 (0.193)& 1.479 (0.134)& 1.221 (0.155)\\
\ion{C}{6}&1032& 1.611 (0.218)& 1.325 (0.095)& 2.039 (0.114)& 1.805 (0.111)& 1.484 (0.091)\\
\ion{O}{6}&1037& 1.305 (0.193)& 1.330 (0.091)& 1.288 (0.086)& 1.538 (0.096)& 1.141 (0.080)\\
\ion{He}{2}+\ion{N}{2}&1085& 0.305 (0.079)& 0.338 (0.109)& 0.257 (0.043)& 0.207 (0.036)& 0.175 (0.052)\\
\ion{N}{5}&1238& 0.373 (0.075)& 0.800 (0.075)& 0.838 (0.088)& 0.720 (0.071)& 0.461 (0.062)\\
\ion{N}{5}&1242& 0.211 (0.066)& 0.514 (0.070)& 0.459 (0.079)& 0.361 (0.068)& 0.343 (0.061)\\
\ion{C}{2}&1335& 0.191 (0.052)& 0.539 (0.084)& 0.225 (0.036)& 0.229 (0.036)& 0.259 (0.077)\\
\ion{O}{5}&1371& 0.150 (0.050)& 0.196 (0.064)& 0.280 (0.039)& 0.227 (0.037)& 0.316 (0.057)\\
\ion{Si}{4}&1400& 0.561 (0.091)& 0.864 (0.105)& 0.564 (0.088)& 0.498 (0.070)& 0.579 (0.086)\\
\ion{O}{4}]&1404& 1.364 (0.296)& 2.461 (0.188)& 1.632 (0.277)& 1.761 (0.173)& 1.980 (0.280)\\
\ion{N}{4}]&1483& ---          & 0.620 (0.059)& 0.375 (0.045)& 0.296 (0.057)& 0.412 (0.113)\\
\ion{N}{4}]&1486& ---          & 0.371        & 0.223        & 0.179        & 0.248        \\
\ion{C}{4}&1547& 1.084 (0.179)& 1.573 (0.132)& 1.048 (0.154)& 1.430 (0.139)& 1.205 (0.143)\\
\ion{C}{4}&1551& 0.811 (0.163)& 1.359 (0.127)& 0.914 (0.150)& 0.995 (0.129)& 0.787 (0.132)\\
\ion{Ne}{4}&1600& 0.193 (0.032)& 0.280 (0.077)& 0.138 (0.041)& 0.157 (0.057)& 0.182 (0.055)\\
\ion{He}{2}&1640& 0.966 (0.168)& 1.261 (0.204)& 1.130 (0.146)& 0.634 (0.091)& 0.932 (0.134)\\
\ion{O}{3}]&1663& 3.188 (0.279)& 4.066 (0.321)& 2.266 (0.214)& 1.998 (0.243)& 3.100 (0.236)\\
\ion{N}{3}]&1749& ---          & 1.155 (0.202)& 0.996 (0.098)& 0.452 (0.111)& 0.573 (0.141)\\
\NeV$\rm^b$&3426& 0.075         & 0.127        & 0.109        & 0.059        & 0.045        \\
\tableline \end{tabular} \\
$\rm^a$Flux in units of $\rm10^{-15} erg~cm^{-2} sec^{-1} arcsec^{-2}$\\
$\rm^b$adapted from Szentgyorgyi et al. (2000)
\end{table}

\begin{table}Table 3: IUE Observation Log\\
\begin{tabular}{lllllrr}
IUE&   Aperture  &RA  &   Decl. & Observation&   Exp & Pos\\
Image& ID &     J2000   &   J2000 &   Date   & Time& Angle\\
Number&  &     hh:mm:ss & dd:mm:ss &        & (ksec)& (deg)\\
\tableline 
SWP26471 &S2M&  20:57:16.33& +31:02:28&  25-Jul-1985 &24.9 &  85 \\
SWP26481 &---&  20:57:11.51& +31:02:47&  26-Jul-1985 &22.2 &  85 \\
SWP28161 &S3M&  20:57:17.84& +31:02:29&  13-Apr-1986 &23.4 & 167 \\
SWP48012 &S5M&  20:57:20.94& +31:02:29&  01-Jul-1993 &21.6 & 108 \\
SWP48013 &S2N&  20:57:16.33& +31:02:43&  01-Jul-1993 &10.8 & 108 \\
SWP48014 &S2S&  20:57:16.32& +31:02:13&  01-Jul-1993 &10.8 & 108 \\
SWP48025 &---&  20:57:21.73& +31:02:58&  02-Jul-1993 &24.0 & 108 \\
SWP48030 &---&  20:57:21.80& +31:00:17&  03-Jul-1993 &25.0 & 106 \\
SWP56022 &---&  20:57:13.52& +31:02:38&  27-Sep-1995 &23.6 &   5 \\
SWP56027 &S1N&  20:57:14.72& +31:02:43&  28-Sep-1995 &23.7 &   4 \\
SWP56031 &S1M&  20:57:14.72& +31:02:28&  29-Sep-1995 &23.1 &   3 \\
SWP56042 &S1S&  20:57:14.72& +31:02:13&  01-Oct-1995 &25.6 &   1 \\
SWP56043 &S1aN &20:57:15.54& +31:02:43&  01-Oct-1995 &27.6 &   1 \\
SWP56044 &S1aM &20:57:15.54& +31:02:28&  02-Oct-1995 &24.9 &   0 \\
SWP56049 &S1aS &20:57:15.54& +31:02:13&  03-Oct-1995 &24.0 &  -1 \\
SWP56050 &S3N  &20:57:17.84& +31:02:43&  03-Oct-1995 &22.0 &  -1 \\
SWP56051 &S3S  &20:57:17.85& +31:02:13&  04-Oct-1995 &24.6 &  -1 \\
SWP56052 &S4N  &20:57:19.43& +31:02:43&  04-Oct-1995 &24.7 &  -2 \\
SWP56053 &S4M  &20:57:19.43& +31:02:29&  05-Oct-1995 &25.2 &  -2 \\
SWP56054 &S4S$\rm ^a$&20:57:19.44& +31:02:14&  05-Oct-1995 &14.4 &  -3 \\
SWP56055 &S4S$\rm ^a$&20:57:19.44& +31:02:14&  06-Oct-1995 & 8.4 &  -3 \\
SWP56056 &S5N  &20:57:21.40& +31:02:44&  06-Oct-1995 &24.0 &  -3 \\
SWP56057 &S5S$\rm ^a$&20:57:21.50& +31:02:14&  06-Oct-1995 &14.4 &  -4 \\
SWP56058 &S5S$\rm ^a$&20:57:21.50& +31:02:14&  07-Oct-1995 &10.2 &  -4 \\
LWP06483 &L5S  &20:57:21.15& +31:02:09&  25-Jul-1985 &22.2 & 85\\    
LWP06493 &L2M  &20:57:16.32& +31:02:29&  26-Jul-1985 &23.3 & 84\\  
LWP31557 &L1N  &20:57:14.72& +31:02:44&  28-Sep-1995 &12.0 &  4\\     
LWP31560 &L1M  &20:57:14.73& +31:02:29&  29-Sep-1995 &18.0 &  3\\ 
\tableline 
\end{tabular} \\
$\rm ^a$coadded exposures 
\end{table}

\begin{table} Table~4: IUE Full Aperture Line Fluxes$\rm ^a$\\
\begin{tabular}{llllllllllll} 
Aperture&\ion{N}{5}&\ion{C}{2}&\ion{O}{4}]$\rm ^b$&\ion{N}{4}]&\ion{C}{4}&\ion{He}{2}&\ion{O}{3}]&\ion{N}{3}]&\ion{Si}{3}]&\ion{C}{3}]&[\ion{C}{2}]$\rm ^c$\\
\AA    & 1241&  1335&  1403&  1486&  1549&  1640&  1665&  1750&  1883&  1909&  2325\\
\tableline 
IUE1N&   0.796& 0.414& 1.879& 0.457& 1.893& 1.106& 1.449& 0.364& 0.682& 2.398& $<$~0.1\\
IUE1M&   0.849& 0.325& 1.616& 0.092& 1.836& 1.154& 1.449& 0.312& 0.811& 2.317& $<$~0.1\\
IUE1S&   0.672& 0.092& 1.483& 0.184& 1.526& 0.450& 0.987& 0.258& 0.634& 1.826&        \\
IUE1aN&  1.101& 0.382& 2.203& 0.725& 2.203& 1.325& 2.584& 0.830& 1.554& 4.815&        \\
IUE1aM&  0.987& 0.411& 1.855& 0.240& 2.517& 0.920& 2.513& 0.382& 0.830& 3.023&        \\
IUE1aS&  0.625& 0.265& 1.325& 0.476& 1.321& 0.687& 1.459& 0.477& 0.486& 1.731&        \\
IUE2N&   1.983& 0.426& 3.809& 0.577& 0.515& 1.821& 3.757& 0.896& 1.755& 7.390&        \\
IUE2M&   1.101& 0.620& 3.657& 0.739& 3.466& 1.507& 3.271& 0.877& 1.893& 9.107& 3.065  \\
IUE2S&   1.035& 0.196& 1.516& 0.149& 1.612& 1.254& 1.950& 0.515& 0.706& 2.064&        \\
IUE3N&   1.049& 0.138& 2.155& 0.400& 2.022& 1.430& 1.712& 0.529& 0.648& 3.800&        \\
IUE3M&   1.654& 0.529& 2.856& 0.496& 2.546& 1.502& 2.312& 0.701& 1.240& 4.420&        \\
IUE3S&   0.606& 0.087& 1.225& 0.214& 1.907& 0.768& 1.283& 0.357& 0.136& 1.726&        \\
IUE4N&   0.934& 0.191& 1.526& 0.150& 2.389& 0.749& 1.597& 0.266& 0.394& 2.317&        \\
IUE4M&   1.359& 0.448& 2.718& 0.663& 2.422& 1.564& 2.207& 0.830& 0.567& 2.746&        \\
IUE4S&   1.201& 0.352& 2.265& 0.606& 2.217& 1.306& 1.755& 0.701& 0.644& 2.365&        \\
IUE5N&   0.154& 0.294& 1.983& 0.577& 1.917& 1.035& 1.998& 0.844& 0.558& 4.410&        \\
IUE5M&   0.615& 0.066& 1.435& 0.301& 1.468& 0.782& 1.073& 0.128& 0.318& 1.697&        \\
IUE5S&   1.101& 0.096& 2.694& 0.710& 1.645& 1.592& 1.797& 0.308& 0.739& 2.589& 0.287  \\
SWP26481&0.277& 0.106& 1.038& 0.086& 1.344& 0.531& 0.565& 0.090& 0.751& 1.349&        \\
SWP48025&0.606& 0.192& 1.511& 0.096& 1.497& 0.801& 1.473& 0.336& 0.815& 1.697&        \\
SWP48030&0.837& 0.069& 0.842& 0.173& 1.158& 0.443& 0.660& 0.124& 0.617& 1.440&        \\
SWP56022&0.501& 0.186& 1.612& 0.359& 1.888& 1.168& 2.212& 0.312& 0.534& 3.676&        \\
Average$\rm ^d$&0.882&0.227&1.869&0.370&1.898&0.944&1.740& 0.424& 0.768& 3.309&        \\
\tableline \end{tabular} \\
$\rm ^a$Flux in units of 10$^{-15}$ erg cm$^{-2}$ sec$^{-1}$ arcsec$^{-2}$ \\
$\rm ^b$includes contributions from \ion{Si}{4} $\lambda$1400\\
$\rm ^c$LWP spectra only\\
$\rm ^d$central XA region only; excludes SWP48030 and SWP26481 \end{table}


\end{document}